\documentclass[pra,twocolumn,superscriptaddress]{revtex4-1}

\usepackage{amssymb}
\usepackage{graphicx}
\usepackage{bbm,amsmath,amssymb}
\usepackage{epsfig}
\usepackage{array}

\usepackage{hyperref}
\hypersetup{
    colorlinks=true,      
   linkcolor=blue,        
   citecolor=red,      
   urlcolor=blue      
}

\begin{document}

\title{From transistor  to trapped-ion computers for quantum chemistry}

\author{M.-H. Yung}
\thanks{These authors contributed equally to this work.}
\address{Department of Chemistry and Chemical Biology, Harvard University, Cambridge MA, 02138, USA}
\address{Center for Quantum Information, Institute for Interdisciplinary Information Sciences, Tsinghua University, Beijing, 100084, P. R. China}

\author{J. Casanova}
\thanks{These authors contributed equally to this work.}
\address{Department of Physical Chemistry, University of the Basque Country UPV/EHU, Apartado 644, 48080 Bilbao, Spain}

\author{A. Mezzacapo}
\address{Department of Physical Chemistry, University of the Basque Country UPV/EHU, Apartado 644, 48080 Bilbao, Spain}

\author{J. McClean}
\address{Department of Chemistry and Chemical Biology, Harvard University, Cambridge MA, 02138, USA}

\author{L. Lamata}
\address{Department of Physical Chemistry, University of the Basque Country UPV/EHU, Apartado 644, 48080 Bilbao, Spain}

\author{A. Aspuru-Guzik}
\email{aspuru@chemistry.harvard.edu}
\address{Department of Chemistry and Chemical Biology, Harvard University, Cambridge MA, 02138, USA}

\author{E. Solano}
\email{enrique.solano@ehu.es}
\address{Department of Physical Chemistry, University of the Basque Country UPV/EHU, Apartado 644, 48080 Bilbao, Spain}
\address{IKERBASQUE, Basque Foundation for Science, Alameda Urquijo 36, 48011 Bilbao,
Spain}

\begin{abstract}
Over the last few decades, quantum chemistry has progressed through the development of computational methods based on modern digital computers. However, these methods can hardly fulfill the exponentially-growing resource requirements when applied to large quantum systems. As pointed out by Feynman, this restriction is intrinsic to all computational models based on classical physics. Recently, the rapid advancement of trapped-ion technologies has opened new possibilities for quantum control and quantum simulations. Here, we present an efficient toolkit that exploits both the internal and motional degrees of freedom of trapped ions for solving problems in quantum chemistry, including molecular electronic structure, molecular dynamics, and vibronic coupling. We focus on applications that go beyond the capacity of classical computers, but may be realizable on state-of-the-art trapped-ion systems. These results allow us to envision a new paradigm of quantum chemistry that shifts from the current transistor to a near-future trapped-ion-based technology.
\end{abstract}

\date{\today}
\maketitle
Quantum chemistry represents one of the most successful applications of quantum mechanics. It provides an excellent platform for understanding matter from atomic to molecular scales, and involves heavy interplay of experimental and theoretical methods. In 1929, shortly after the completion of the basic structure of the quantum theory, Dirac speculated~\cite{Dirac1929} that the fundamental laws for chemistry were completely known, but the application of the fundamental laws led to equations that were too complex to be solved. About ninety years later, with the help of transistor-based digital computers, the development of quantum chemistry continues to flourish, and many powerful methods, such as Hartree-Fock, configuration interaction, density functional theory, coupled-cluster, and quantum Monte Carlo, have been developed to tackle the complex equations of quantum chemistry (see e.g.~\cite{Love2012} for a historical review). However, as the system size scales up, all of the methods known so far suffer from limitations that make them fail to maintain accuracy with a finite amount of resources~\cite{Head-Gordon2008}. In other words, quantum chemistry remains a hard problem to be solved by the current computer technology. 

As envisioned by Feynman~\cite{Feynman1982}, one should be able to efficiently solve problems of quantum systems with a quantum computer. Instead of solving the complex equations, this approach, known as {\it quantum simulation} (see the recent reviews in Refs.~\cite{Kassal2011, Yungintro2012,Aspuru2012}),  aims to solve the problems by simulating target systems with another controllable quantum system, or qubits. Indeed, simulating many-body systems beyond classical resources will be a cornerstone of quantum computers. Quantum simulation is a very active field of study and various methods have been developed. Quantum simulation methods have been proposed for preparing specific states such as ground~\cite{Abrams1999,Aspuru-Guzik2005c,Poulin2009c,Lanyon2010,Li2011,Xu2012} and thermal states~\cite{Lidar1997,Poulin2009a,Yung2010,Bilgin2010,Temme2009,Zhang2012,Yung17012012}, simulating time evolution~\cite{Lloyd1996,Zalka1998b,Wu2002,Kassal2008,Lanyon2011a,Childs2011,Casanova12}, and the measurement of physical observables~\cite{Lidar1999,Master2003,Kassal2009,Wocjan2009a}.

\begin{figure*}[t!]
\begin{center}
\includegraphics [width= 2 \columnwidth]{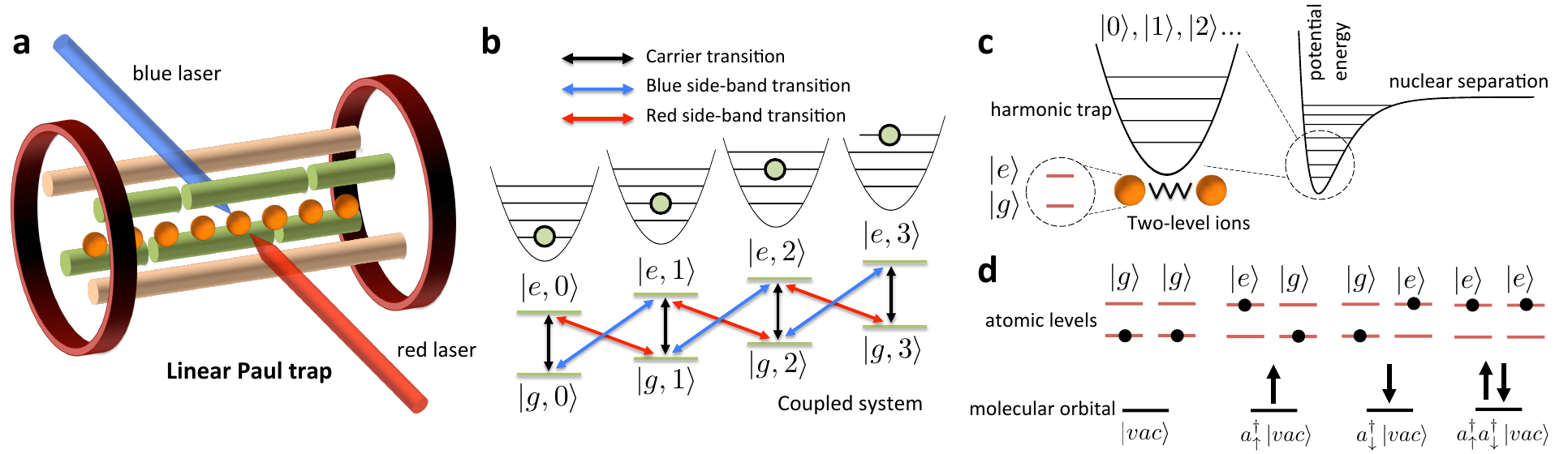}
\end{center}
\caption{Simulating quantum chemistry with trapped ions. (a) Scheme of a trapped-ion setup for quantum simulation, which contains a linear chain of trapped ions confined by a harmonic potential, and external lasers that couple the motional and internal degrees of freedom. (b) Transitions between internal and motional degrees of freedom of the ions in the trap. (c) The normal modes of the trapped ions can simulate the vibrational degrees of freedom of molecules. (d) The internal states of two ions can simulate all four possible configurations of a molecular orbital.} \label{fig:trappedions}
\end{figure*}

Trapped-ion systems (see Fig.~\ref{fig:trappedions}) are currently one of the most sophisticated technologies developed for quantum information processing~\cite{Haeffner2008}. These systems  offer an unprecedented level of quantum control, which opens new possibilities for obtaining physico-chemical information about quantum chemical problems.   The power of trapped ions for quantum simulation is manifested by the high-precision control over both the internal degrees of freedom of the individual ions and the phonon degrees of freedom of the collective motions of the trapped ions, and the high-fidelity initialization and measurement~\cite{Haeffner2008,Leibfried2003}. Up to 100 quantum logic gates have been realized for six qubits with trapped ions~\cite{Lanyon2011a}, and quantum simulators involving 300 ions have been demonstrated~\cite{Britton2012}. 

\begin{figure*}[t]
\begin{center}
\includegraphics [width= 2 \columnwidth]{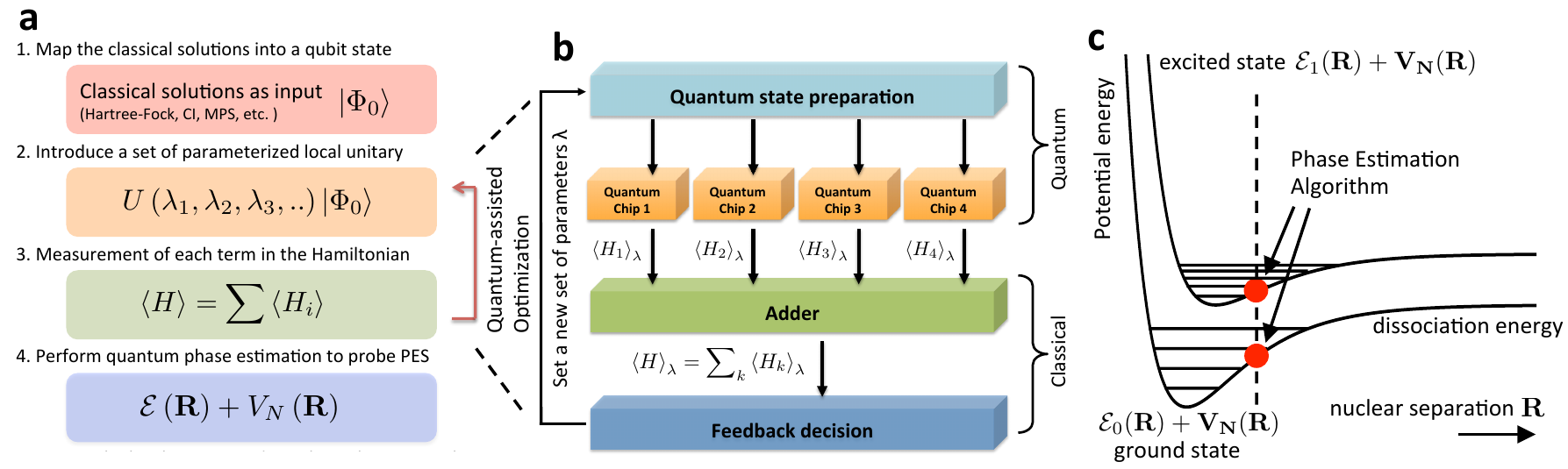}
\end{center}
\caption{Outline of the quantum-assisted optimization method. (a) The key steps for quantum assisted optimization, which starts from classical solutions. For each new set of parameters $\lambda$'s, determined by a classical optimization algorithm, the expectation value $\left\langle H \right\rangle $ is calculated. The potential energy surface is then obtained by quantum phase estimation. (b) Quantum measurements are performed for the individual terms in $H$, and the sum is obtained classically. (c) The same procedure is applied for each nuclear configuration $\bf R$ to probe the energy surface. }
\label{fig:QOptimize}
\end{figure*}

In this work, we present an efficient toolkit for solving quantum chemistry problems based on the state-of-the-art trapped-ion technologies. The toolkit comprises two components {\it i}) First, we present a hybrid quantum-classical variational optimization method, called quantum-assisted optimization, for approximating both ground-state energies and the ground-state eigenvectors for electronic problems. The optimized eigenvector can then be taken as an input for the phase estimation algorithm to project out the exact eigenstates and hence the potential-energy surfaces (see Fig.~\ref{fig:QOptimize}). Furthermore, we extend the application of the unitary coupled-cluster method~\cite{Taube2006}. This allows for the application of a method developed for classical numerical computations in the quantum domain. {\it ii}) The second main component of our toolkit is the optimized use of trapped-ion phonon degrees of freedom not only for quantum-gate construction, but also for simulating molecular vibrations, representing a mixed digital-analog quantum simulation. The phonon degrees of freedom in trapped-ion systems provide a natural platform for addressing spin-boson or fermion-boson-type problems through quantum simulation~\cite{Lamata2007,Gerritsma10,Casanova11,Mueller2011,Casanova12,Mezzacapo12}. It is noteworthy to mention that, contrary to the continuous of modes required for full-fledged quantum field theories, quantum simulations of quantum chemistry problems could reach realistic conditions for finite bosonic and fermionic mode numbers. Consequently, trapped ions can be exploited to solve dynamical problems involving linearly or non-linearly coupled oscillators, e.g., spin-boson models~\cite{Leggett1987,Mostame2012}, that are difficult to solve either analytically or numerically with a classical computer. Furthermore, we have also developed a novel protocol to measure correlation functions of observables in trapped ions that will be crucial for the quantum simulation of quantum chemistry.

\section*{Results and discussion}\label{qchybrid}

\subsection*{Trapped ions for quantum chemistry}
Quantum chemistry deals with the many-body problem involving electrons and nuclei. Thus, it is very well suited for being simulated with trapped-ion systems, as we will show below. The full quantum chemistry Hamiltonian, $H = {T_e} + {V_e} + {T_N} + {V_N} + {V_{eN}}$, is a sum of the kinetic energies of the electrons ${T_e} \equiv  - {\textstyle{{{\hbar ^2}} \over {2m}}}\sum\nolimits_i {\nabla _{e,i}^2}$ and nuclei ${T_N} \equiv  - \sum\nolimits_i {{\textstyle{{{\hbar ^2}} \over {2{M_i}}}}\nabla _{N,i}^2}$, and the electron-electron ${V_e} \equiv \sum\nolimits_{j > i} {{e^2}} /\left| {{{\bf r}_i} - {{\bf r}_j}} \right|$, nuclei-nuclei ${V_N} \equiv \sum\nolimits_{j > i} {{Z_i}{Z_j}{e^2}} /\left| {{{\bf R}_i} - {{\bf R}_j}} \right|$, and electron-nuclei ${V_{eN}} \equiv  - \sum\nolimits_{i,j} {{Z_j}{e^2}} /\left| {{{\bf r}_i} - {{\bf R}_j}} \right|$ potential energies, where ${\bf r}$ and ${\bf R}$ respectively refer to the electronic and nuclear coordinates. 

In many cases, it is more convenient to work on the second-quantization representation for quantum chemistry. The advantage is that one can choose a good fermionic basis set of molecular orbitals, $\left| p \right\rangle  = c_p^\dagger \left| {vac} \right\rangle$, which can compactly capture the low-energy sector of the chemical system. This kind of second quantized fermionic Hamiltonians are efficiently simulatable in trapped ions~\cite{Casanova12}. To be more specific, we will choose first $M>N$ orbitals for an $N$-electron system. Denote ${\phi _p}\left( {\bf r} \right) \equiv \left\langle {\bf r} \right|  {p} \rangle$ as the single-particle wavefunction corresponding to mode $p$. The electronic part, ${H_e} ({\bf R}) \equiv {T_e} + {V_{eN}}\left( {\bf R} \right) + {V_e}$, of the Hamiltonian $H$ can be expressed as follows: 
\begin{equation}\label{h_pq_h_pqrs}
{H_e}({\bf R}) = \sum\limits_{pq} {{h_{pq}}c_p^\dagger {c_q}}  + \frac{1}{2}\sum\limits_{pqrs} {{h_{pqrs}}c_p^\dagger c_q^ \dagger {c_r}{c_s}},
\end{equation}
where $h_{pq}$ is obtained from the single-electron integral ${h_{pq}} \equiv  - \int {d{\bf r}} \phi _p^*\left( {\bf r} \right)\left( {{T_e} + {V_{eN}}} \right){\phi _q}\left( {\bf r} \right)$, and $h_{pqrs}$ comes from the electron-electron Coulomb interaction, ${h_{pqrs}} \equiv \int {d{{\bf r}_1}d{{\bf r}_2}} \phi _p^*\left( {{{\bf r}_1}} \right)\phi _q^*\left( {{{\bf r}_2}} \right){{V_e}\left( {\left| {{{\bf r}_1} - {{\bf r}_2}} \right|} \right)}{\phi _r}\left( {{{\bf r}_2}} \right){\phi _s}\left( {{{\bf r}_1}} \right)$. We note that the total number of terms in $H_e$ is $O(M^4)$; typically $M$ is of the same order as $N$. Therefore, the number of terms in $H_e$ scales polynomially in $N$, and the integrals $\left\{ {{h_{pq}},{h_{pqrs}}} \right\}$ can be numerically calculated by a classical computer with polynomial resources~\cite{Aspuru-Guzik2005c}. 

To implement the dynamics associated with the electronic Hamiltonian in Eq.~(\ref{h_pq_h_pqrs}) with a trapped-ion quantum simulator, one should take into account the fermionic nature of the operators $c_p$ and $c_q^\dagger$. We invoke the Jordan-Wigner transformation (JWT), which is a method for mapping the occupation representation to the spin (or qubit) representation~\cite{Ortiz2001}. Specifically, for each fermionic mode $p$, an unoccupied state ${\left| 0 \right\rangle _p}$ is represented by the spin-down state ${\left|  \downarrow  \right\rangle _p}$, and an occupied state ${\left| 1 \right\rangle _p}$ is represented by the spin-up state ${\left|  \uparrow  \right\rangle _p}$. The exchange symmetry is enforced by the Jordan-Wigner transformation: $c_p^ \dagger  = ( {\prod\nolimits_{m < p} {\sigma _m^z} } ) \sigma _p^ +$ and ${c_p} = ( {\prod\nolimits_{m < p} {\sigma _m^z} } )\sigma _p^ -$, where ${\sigma ^ \pm } \equiv \left( {{\sigma ^x} \pm i{\sigma ^y}} \right)/2$. Consequently, the electronic Hamiltonian in Eq.~(\ref{h_pq_h_pqrs}) becomes highly nonlocal in terms of the Pauli operators $\left\{ {{\sigma ^x},{\sigma ^y},{\sigma ^z}} \right\}$, i.e.,
\begin{equation}\label{He_JWT}
{H_e}  \mathrel{\mathop{\kern0pt\longrightarrow} \limits_{{{\rm JWT}}}}  \sum\limits_{i,j,k...\in \left\{ {x,y,z} \right\}} {{g_{ijk...}}\left( {\sigma _1^i \otimes \sigma _2^j \otimes \sigma _3^k...} \right)} \quad.
\end{equation} 
Nevertheless, the simulation can still be made efficient with trapped ions, as we shall discuss below.

In trapped-ion physics two metastable internal levels of an ion are typically employed as a qubit. Ions can be confined either in Penning traps or radio frequency Paul traps~\cite{Leibfried2003}, and cooled down to form crystals. Through sideband cooling the ions motional degrees of freedom can reach the ground state of the quantum Harmonic oscillator, that can be used as a quantum bus to perform gates among the different ions. Using resonance fluorescence with a cycling transition quantum non demolition measurements of the qubit can be performed. The fidelities of state preparation, single- and two-qubit gates, and detection, are all above 99\%~\cite{Haeffner2008}.

The basic interaction of a two-level trapped ion with a single-mode laser is given by~\cite{Haeffner2008}, $H=\hbar\Omega\sigma_+e^{-i(\Delta t-\phi)}\exp(i\eta[a e^{-i\omega_t t}+a^\dag e^{i\omega_t t}])+{\rm H.c.}$, where $\sigma_\pm$ are the atomic raising and lowering operators, $a$ ($a^\dag$) is the annihilation (creation) operator of the considered motional mode, and $\Omega$ is the Rabi frequency associated to the laser strength. $\eta=k z_0$ is the Lamb-Dicke parameter, with $k$ the wave vector of the laser and $z_0=\sqrt{\hbar/(2m\omega_t)}$ the ground state width of the motional mode. $\phi$ is a controllable laser phase and $\Delta$ the laser-atom detuning. 

In the Lamb-Dicke regime where $\eta\sqrt{\langle (a+a^\dag)^2\rangle}\ll 1$, the basic interaction of a two-level trapped ion with a laser can be rewritten as $H =  \hbar\Omega[\sigma_+e^{-i(\Delta t-\phi)}+i\eta\sigma_+e^{-i(\Delta t-\phi)}(a e^{-i\omega_t t}+a^\dag e^{i\omega_t t})$ + {H.c.}

By adjusting the laser detuning $\Delta$, one can generate the three basic ion-phonon interactions, namely: the carrier interaction ($\Delta=0$), $H_c=\hbar\Omega(\sigma_+e^{i\phi}+\sigma_-e^{-i\phi})$,
the red sideband interaction, ($\Delta=-\omega_t$), $H_r=i\hbar\eta\Omega(\sigma_+a e^{i\phi}-\sigma_- a^\dag e^{-i\phi})$, and the blue sideband interaction, ($\Delta=\omega_t$), $H_b=i\hbar\eta\Omega(\sigma_+a^\dag e^{i\phi}-\sigma_- a e^{-i\phi})$. By combining detuned red and blue sideband interactions, one obtains the M\o lmer-S\o rensen gate~\cite{MolmerSorensen}, which is the basic building block for our methods. With combinations of this kind of gates, one can obtain dynamics as the associated one to $H_e$ in Eq. (\ref{He_JWT}), that will allow one to simulate arbitrary quantum chemistry systems.

\subsection*{Quantum-assisted optimization}
Quantum-assisted optimization~\cite{Peruzzo2013} (see also Fig.~\ref{fig:QOptimize}) for obtaining ground-state energies aims to optimize the use of quantum coherence by breaking down the quantum simulation through the use of both quantum and classical processors; the quantum processor is strategically employed for expensive tasks only.

To be more specific, the first step of quantum-assisted optimization is to prepare a set of quantum states $\{ \left| {{\psi _\lambda }} \right\rangle \}$ that are characterized by a set of parameters $\{ \lambda \}$.  After the state is prepared, the expectation value $ E_\lambda \equiv \left\langle {{\psi _\lambda }} \right|H\left| {{\psi _\lambda }} \right\rangle$ of the Hamiltonian $H$ will be measured directly, without any quantum evolution in between. Practically, the quantum resources for the measurements can be significantly reduced when we divide the measurement of the Hamiltonian $H = \sum\nolimits_i {{H_i}} $ into a polynomial number of small pieces ${\left\langle {{H_i}} \right\rangle }$ (cf Eq.~(\ref{He_JWT})). These measurements can be performed in a parallel fashion, and no quantum coherence is needed to maintain between the measurements (see Fig.~\ref{fig:QOptimize}a and \ref{fig:QOptimize}b). Then, once a data point of $E_\lambda$ is obtained, the whole procedure is repeated for a new state $\{ \left| {{\psi _\lambda' }} \right\rangle \}$ with another set of parameters $\{ \lambda' \}$. The choice of the new parameters is determined by a classical optimization algorithm that aims to minimize $E_\lambda$ (see Methods). The optimization procedure is terminated after the value of $E_\lambda$ converges to some fixed value. 

Finally, for electronic Hamiltonians $H_e ({\bf R})$, the optimized state can then be sent to a quantum circuit of phase estimation algorithm to produce a set of data point for some $\bf R$ on the potential energy surfaces (Fig. ~\ref{fig:QOptimize}c shows the 1D case). After locating the local minima of the ground and excited states, vibronic coupling for the electronic structure can be further studied (see Supplementary Material).  

The performance of quantum-assisted optimization depends crucially on (a) the choice of the variational states, and (b) efficient measurement methods. We found that the unitary coupled-cluster (UCC) states~\cite{Taube2006} are particularly suitable for being the input state for quantum-assisted optimization, where each quantum state $\left| {{\psi _\lambda }} \right\rangle$ can be prepared efficiently with a digital quantum circuit and with trapped ions. Furthermore, efficient measurement methods for $H_e$ are also available for trapped ion systems. We shall discuss these results in detail in the following sections.

\subsection*{Unitary coupled-cluster (UCC) ansatz \label{SectCCUCC}}
 The unitary coupled-cluster (UCC) ansatz~\cite{Taube2006} assumes electronic states $\left| \psi  \right\rangle$ have the following form, $\left| \psi  \right\rangle  = {e^{T - {T^\dagger }}}\left| \Phi  \right\rangle$,
where $\left| \Phi  \right\rangle$ is a reference state, which can be, e.g., a Slater determinant constructed from Hartree-Fock molecular orbitals. The particle-hole excitation operator, or cluster operator $T$, creates a linear combination of excited Slater determinants from $\left| \Phi  \right\rangle$. Usually, $T$ is divided into subgroups based on the particle-hole rank. More precisely, $T = {T_1} + {T_2} + {T_3} + ... + {T_N}$ for an $N$-electron system, where ${T_1} = \sum\nolimits_{i,a} { {t_i^ac_a^\dagger {c_i}} }$, ${T_2} = \sum\nolimits_{i,j,a,b} { {t_{ij}^{ab}c_a^ \dagger c_b^\dagger {c_j}{c_i}} }$, and so on.

Here $c_a^\dagger$ creates an electron in the orbital $a$. The indices $a,b$ label unoccupied orbitals in the reference state $\left| \Phi  \right\rangle$, and  $i,j$ label occupied orbitals. The energy obtained from UCC, namely $E = \left\langle \Phi  \right|{e^{{T^\dagger } - T}}H{e^{T - {T^\dagger}}}\left| \Phi  \right\rangle$ is a variational upper bound of the exact ground-state energy. 

The key challenge for implementing UCC on a classical computer is that the computational resource grows exponentially. It is because, in principle, one has to expand the expression $\tilde H \equiv {e^{{T^\dagger } - T}}H{e^{T - {T^\dagger }}}$ into an infinity series, 
using the Baker-Campbell-Hausdorff expansion. Naturally, one has to rely on approximate methods~\cite{Kutzelnigg1991,Taube2006} to truncate the series and keep track of finite numbers of terms. Therefore, in order to make good approximations by perturbative methods, i.e., assuming $T$ is small, one implicitly assumes that the reference state $\left| \Phi  \right\rangle$ is a good solution to the problem. However, in many cases, such an assumption is not valid and the use of approximate UCC breaks down. We explain below how implementing UCC on a trapped-ion quantum computer can overcome this problem.

\begin{table*}[t!]
\caption{Using trapped ions to simulate quantum chemistry}
\begin{ruledtabular}
\begin{tabular}{ p{4.3cm}p{6.3cm}p{6.3cm}} 
& {\bf  Simulating Quantum Chemistry} & {\bf  Implementation with Trapped Ions} \\
& &  \\
{ Hamiltonian transformation:} & The fermionic (electronic) Hamiltonian $H_e$ is transformed into a spin Hamiltonian through the Jordan-Wigner transformation. & The spin degrees of freedom in $H_e$ are represented by the internal degrees of freedom of the trapped ions.  \\
& & \\
& \multicolumn{2}{c}{${H_e} \ \to \sum\limits_{i,j,k, \cdots \in \left\{ {x,y,z} \right\}} {{g_{ijk \cdots}}\left( {\sigma _1^i \otimes \sigma _2^j \otimes \sigma _3^k \cdots } \right)}  \equiv \sum\limits_{l=1}^m {{H_l}} $} \\
& & \\
{Simulation of time evolution:} & The time evolution operator ${e^{ - i{H_e}t}}$ is split into $n$ small-time ($t/n$) pieces ${e^{ - i{H_l}t/n}}$ through the Suzuki-Trotter expansion. & Each individual term ${e^{ - i{H_l}t/n}}$ can be simulated with trapped ions through the use of M\o lmer-S\o rensen gates $U_{MS}$. Explicitly, ${e^{ - i{H_l}t/n}} = {U_{MS}}\left( {{\textstyle{{ - \pi } \over 2}},0} \right){U_{{\sigma _z}}}\left( \phi  \right){U_{MS}}\left( {{\textstyle{\pi  \over 2}},0} \right)$.\\
& & \\
 & \multicolumn{2}{c}{${e^{ - i\sum\nolimits_{l=1}^m {{H_{l}}} t}} \approx {( {{e^{ - i{H_1}t/n}}{e^{ - i{H_2}t/n}} \cdots {e^{ - i{H_m}t/n}}} )^n}$} \\
 & & \\
{Measuring eigenvalues:} & The eigenvalues of the Hamiltonian can be obtained through the phase estimation algorithm. Good trial states can be obtained through classical computing, or the unitary coupled-cluster method. & The phase estimation algorithm can be implemented through the simulation of controlled time evolutions. \\
 & & \\
 {Obtaining average energy:} & The average energy $\left\langle {{H_e}} \right\rangle$ of the Hamiltonian can be obtained through the sum of the individual terms $\left\langle {{H_l}} \right\rangle$, which reduces to the measurement of products of Pauli matrices. & For any prepared state $\left| \psi  \right\rangle$, average values of the products of Pauli matrices ${J_{ijk...}} \equiv \sigma _1^i \otimes \sigma _2^j \otimes \sigma _3^k \cdots$ can be measured by first applying the pseudo time evolution operator ${e^{ - i\left( {\pi /4} \right){J_{ijk \cdots }}}}$ to $\left| \psi  \right\rangle$ and then measuring $\left\langle {\sigma _1^z} \right\rangle$. \\
 & & \\
{Molecular vibrations:} & The inclusion of vibrational degrees of freedom is necessary for corrections on the Born-Oppenheimer picture in the electronic structure of molecules. & The vibrational degrees of freedom are represented by the quantized vibrational motion of the trapped ions.
\end{tabular}
\end{ruledtabular}
\label{table_complexity}
\end{table*}

\subsection*{Implementation of UCC through time evolution}
We can generate the UCC state by simulating a pseudo time evolution through Suzuki-Trotter expansion on the evolution operator ${e^{T - {T^\dagger }}}$~\cite{Lloyd1996}. To proceed, we consider an $N$-electron system with $M$, where $M>N$, molecular orbitals (including spins). We need totally $M$ qubits; the reference state is the Hartree-Fock state where $N$ orbitals are filled, and $M-N$ orbitals are empty, i.e, $\left| \Phi  \right\rangle  = | { {000..0} {111..1}} \rangle$. We also define an effective Hamiltonian $K \equiv i\left( {T - {T^\dagger }} \right)$, which means that we should prepare the state ${e^{ - iK}}\left| \Phi  \right\rangle.$

We decompose $K$ into subgroups $K = {K_1} + {K_2} + {K_3} + ...+{K_P}$, where $P \le N$, and ${K_i} \equiv i( {{T_i} - T_i^ \dagger})$. We now write ${e^{ - iK}} = {\left( {{e^{ - iK\delta }}} \right)^{1/\delta }}$ for some dimensionless constant $\delta$. For small $\delta$, we have ${e^{ - iK\delta }} \approx {e^{ - i{K_P}\delta }}...{e^{ - i{K_2}\delta }}{e^{ - i{K_1}\delta }}$. Since each ${{K_j}}$ contains ${N^j}{\left( {M - N} \right)^j}$ terms of the creation $c^\dagger$ and annihilation $c$ operators, we will need to individually simulate each term separately, e.g., ${e^{ - i\left( {tc_a^\dagger {c_i} - {t^*}c_i^\dagger {a_a}} \right)}}$ and ${e^{ - i\left( {tc_a^\dagger c_b^\dagger {c_j}{c_i} - {t^*}c_i^ \dagger c_j^ \dagger {c_b}{c_a}} \right)}}$, which can be implemented by transforming into spin operators through Jordan-Wigner transformation. The time evolution for each term can be simulated with a quantum circuit involving many nonlocal controlled gates, which can be efficiently implemented with trapped ions as we shall see below.

 \subsection*{Implementation of UCC and simulation of time evolution with trapped-ions\label{TimeEvol}}

Our protocol for implementing the UCC ansatz requires the simulation of the small-time $t/n$ evolution of non-local product of Pauli matrices of the form: $e^{-iH_l t /n }$, where ${H_l} = {g_l}\sigma _1^i\sigma _2^j\sigma _3^k \cdots $ for $i,j,k \in \left\{ {x,y,z} \right\}$. Note that for any $N$-spin interaction,  the $e^{-iH_l t /n }$ terms are equivalent to ${e^{i\phi \sigma _1^z\sigma _2^x\sigma _3^x \cdots \sigma _N^x}}$ through local spin rotations, which are simple to implement on trapped ions. Such a non-local operator can be implemented using the multi-particle M\o lmer-S\o rensen gate~\cite{Mueller2011, Casanova12}: $U_{\rm MS}(\theta, \varphi) \equiv \exp{\left[ -i\theta(\cos\varphi S_x + \sin\varphi S_{y})^2/4   \right]}$, where $S_{x, y} \equiv \sum_{i} \sigma^{x, y}_i$ is a collective spin operator. Explicitly,
\begin{equation}\label{product_MS}
{e^{i\phi \sigma _1^z\sigma _2^x\sigma _3^x \cdots \sigma _N^x}} = {U_{MS}}\left( {{\textstyle{{ - \pi } \over 2}},0} \right){R_N (\phi)} {U_{MS}}\left( {{\textstyle{\pi  \over 2}},0} \right) \quad.
\end{equation} 
Here ${R_N (\phi)}$ is defined as follows: for any $m \in \mathbb{N}$, $R_N (\phi) = {e^{ \pm i\phi \sigma _1^z}}$ for $N=4m\pm1$, and (ii) ${R_N}\left( \phi  \right) = {e^{i\phi \sigma _1^y}}$ for $N=4m$, and (iii) ${R_N}\left( \phi  \right) = {e^{-i\phi \sigma _1^y}}$ for $N=4m-2$.

It is remarkable that the standard quantum-circuit treatment (e.g. see Ref.~\cite{Whitfield2011}) for implementing each $e^{-iH_l t /n }$ involves as many as $2N$ two-qubit gates for simulating $N$ fermionic modes; in our protocol one needs only two M\o lmer-S\o rensen gates, which are straightforwardly implementable with current trapped-ion technology. Furthermore, the local rotation $R_N(\phi)$ can also include motional degrees of freedom of the ions for simulating arbitrary fermionic Hamiltonians coupled linearly to bosonic operators $a_k$ and $a_k^\dag$. 

\subsection*{Measurement of arbitrarily-nonlocal spin operators}\label{sec:measure_w_ions}
For any given state $|\psi\rangle$, we show how to encode expectation value of products of Pauli matrices $\langle \sigma _1^i \otimes \sigma _2^j \otimes \sigma _3^k \otimes  \cdots \rangle  \equiv \left\langle \psi  \right|\sigma _1^i \otimes \sigma _2^j \otimes \sigma _3^k \otimes  \cdots \left| \psi  \right\rangle $, where $i, j, k  \in \{ x, y, z\}$, onto an expectation value of a single qubit. The idea is to first apply the unitary evolution of the form: $e^{-i \theta (\sigma^i_1\otimes \sigma^j_2\otimes \cdots )}$, which as we have seen (cf Eq.~\ref{product_MS}) can be generated by trapped ions efficiently, to the state $|\psi\rangle$ before the measurement. For example, defining $\left| {{\psi _\theta }} \right\rangle  \equiv {e^{ - i\theta \left( {\sigma _1^x \otimes \sigma _2^x \otimes  \cdots } \right)}}\left| \psi  \right\rangle$, we have the relation 
\begin{equation}\label{mea_nonlocal_spins}
\left\langle {{\psi _\theta }} \right|\sigma _1^z\left| {{\psi _\theta }} \right\rangle  = \cos \theta \left\langle {\sigma _1^z} \right\rangle  + \sin \theta \left\langle {\sigma _1^y \otimes \sigma _2^x \otimes  \cdots } \right\rangle  \quad,
\end{equation}
which equals $\langle\psi| (\sigma^y_1\otimes \sigma^x_2\otimes ...) |\psi\rangle$ for $\theta  = \pi /4$. Note that the application of this method requires the measurement of one qubit only,  making this technique especially suited for trapped ion systems where the fidelity of the measurement of one qubit is $99.99\%$~\cite{Myerson2008}.

This method can be further extended to include bosonic operators in the resulting expectation values. For example, re-define $\left| {{\psi _\theta }} \right\rangle  \equiv {e^{ - i\theta (\sigma _1^i \otimes \sigma _2^j \otimes  \cdots ) \otimes \left( {a + {a^\dag }} \right)}}\left| \psi  \right\rangle$ and consider $\theta  \to \theta \left( {a + {a^\dag }} \right)$ in Eq.~(\ref{mea_nonlocal_spins}). We can obtain the desired correlation through the derivative of the single-qubit measurement:  ${\left. {{\partial _\theta }\left\langle {{\psi _\theta }} \right|\sigma _1^z\left| {{\psi _\theta }} \right\rangle } \right|_{\theta  = 0}} =  - 2 \langle {\left( {\sigma _1^y \otimes \sigma _2^x \otimes  \cdots } \right) ( {a + {a^\dag }} )} \rangle$. Note that the evolution operator of the form ${e^{ - i\theta (\sigma _1^i \otimes \sigma _2^j \otimes  \cdots ) \otimes ( {a + {a^\dag }} )}}$ can be generated by replacing the local operation $R_N (\phi)$ in Eq.~\ref{product_MS} with ${e^{ \pm i\phi \sigma _1^i ( {a + {a^\dag }} )}}$. This technique allows us to obtain a diverse range of correlations between bosonic and internal degrees of freedom.  

\subsection*{Probing potential energy surfaces}

In the Born-Oppenheimer (BO) picture, the potential energy surface ${\cal E}_k \left( {\bf R} \right) + {V_N}\left( {\bf R} \right)$ associated with each electronic eigenstate $\left| {{\phi _k}} \right\rangle$ is obtained by scanning the eigenvalues ${\cal E}_k \left( {\bf R} \right)$ for each configurations of the nuclear coordinates~$\{ \bf R \}$. Of course, we can apply the standard quantum phase estimation algorithm~\cite{Kaye2007} that allows us to extract the eigenvalues. However, this can require many ancilla qubits. In fact, locating these eigenvalues can be achieved by the phase estimation method utilizing one extra ancilla qubit~\cite{Li2011}. 

This method works as follows: suppose we are given a certain quantum state~$\left| \psi  \right\rangle$ (which may be obtained from classical solutions with quantum-assisted optimization) and an electronic Hamiltonian $H_e ({\bf R})$ (cf.~Eq.~(\ref{h_pq_h_pqrs})). Expanding the input state, $\left| \psi  \right\rangle  = \sum\nolimits_k {{\alpha_k}} \left| {{\phi _k}} \right\rangle$, by the eigenstate vectors $\left| {{\phi _k}} \right\rangle$ of $H_e ({\bf R})$, where ${H_e}\left( {\bf R} \right)\left| {{\phi _k}} \right\rangle  = {{\cal E}_k}\left( {\bf R} \right)\left| {{\phi _k}} \right\rangle$, then for the input state $\left| 0 \right\rangle \left| \psi  \right\rangle$, the quantum circuit of the quantum phase estimation produces the following output state, $\left( {1/\sqrt 2 } \right)\sum\nolimits_k {{\alpha_k}\left( {\left| 0 \right\rangle  + {e^{ - i{\omega _k}t}}\left| 1 \right\rangle } \right)\left| {{\phi _k}} \right\rangle }$, where ${\omega _k} = {{\mathcal{E}}_k}/\hbar$. The corresponding reduced density matrix,
\begin{equation}
\frac{1}{2}\left( {\begin{array}{*{20}{c}}
1&{\sum\nolimits_k {{{\left| {{\alpha_k}} \right|}^2}{e^{i{\omega _k}t}}} }\\
{\sum\nolimits_k {{{\left| {{\alpha_k}} \right|}^2}{e^{ - i{\omega _k}t}}} }&1
\end{array}} \right),
\end{equation}
of the ancilla qubit contains the information about the weight (amplitude-square) ${{{\left| {{\alpha_k}} \right|}^2}}$ of the eigenvectors $\left| {{\phi _k}} \right\rangle$ in $\left| \psi  \right\rangle$ and the associated eigenvalues $\omega_k$ in the off-diagonal matrix elements. All ${{{\left| {{\alpha_k}} \right|}^2}}$'s and $\omega_k$'s can be extracted by repeating the quantum circuit for a range of values of $t$ and performing a (classical) Fourier transform to the measurement results. The potential energy surface is obtained by repeating the procedure for different values of the nuclear coordinates $\{\bf R\}$.

 \begin{figure*}[t!]
\begin{center}
\includegraphics [width= 2.05 \columnwidth]{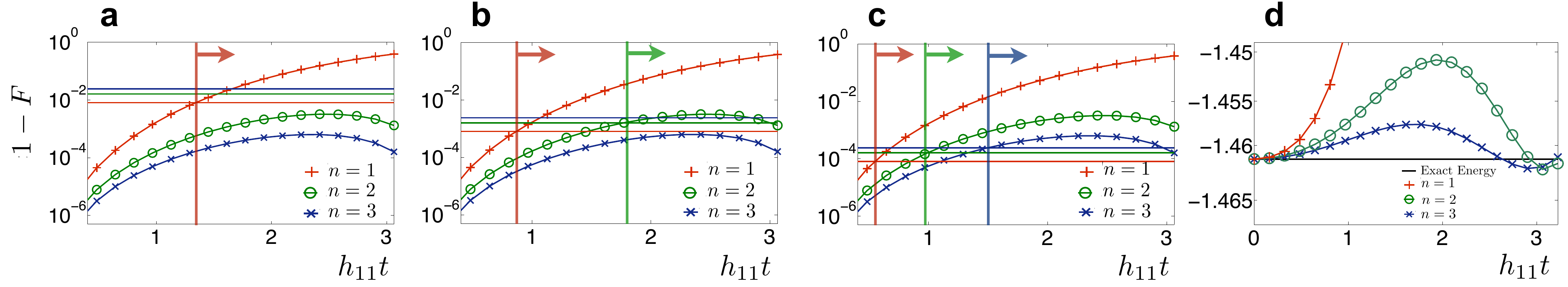}
\end{center}
\caption{(Color online). Digital error $1-F$ (curves) along with the accumulated gate error (horizontal lines)  versus time in $h_{11}$ energy units, for $n=1,2,3$ Trotter steps in each plot, considering a protocol with an error per Trotter step of $\epsilon=10^{-3}$ (a), $\epsilon=10^{-4}$ (b) and $\epsilon=10^{-5}$ (c). The initial state considered is $|\!\!\uparrow\uparrow\downarrow\downarrow\rangle$, in the qubit representation of the Hartree-Fock
state in a molecular orbital basis with one electron on the first and second orbital. Vertical lines and arrows define the time domain in which the dominant part of the error is due to the digital approximation. d) Energy of the system, in $h_{11}$ units, for the initial state $|\uparrow\uparrow\downarrow\downarrow\rangle$ for the exact dynamics, versus the digitized one. For a protocol with three Trotter steps the energy is recovered up to a negligible error.}\label{FidelityLoss} 
\end{figure*}

\subsection*{Numerical investigation}

 In order to show the feasibility of our protocol, we can estimate the trapped-ion resources needed to simulate, e.g., the prototypical electronic Hamiltonian ${H_e} = \sum {{h_{pq}}a_p^\dagger {a_q}}  + (1/2)\sum {{h_{pqrs}}a_p^\dagger a_q^ \dagger {a_r}{a_s}}$ as described in Eq.~(\ref{h_pq_h_pqrs}), for the specific case of the H$_2$ molecule in a minimal STO-3G basis.  This is a two-electron system represented in a basis of four spin-orbitals.  The hydrogen atoms were separated by 0.75 \AA, near the equilibrium bond distance of the molecule. The Hamiltonian is made up of $12$ terms, that include $4$ local ion operations and $8$ non-local interactions. Each of the non-local terms can be done as a combination of two M\o lmer-S\o rensen (MS) gates and local rotations, as described in Table~\ref{table_complexity}. Therefore, to implement the dynamics, one needs $16$ MS gates per Trotter step and a certain number of local rotations upon the ions. Since $\pi/2$ MS gates can be done in $\sim 50\mu$s, and local rotations can be performed in negligible times ($\sim1\mu$s)~\cite{Lanyon2011a,Haeffner2008}, the total simulation time can be assumed of about $800$ $\mu$s for the $n=1$ protocol, $1.6$ ms and $2.4$ ms for the $n=2$ and $n=3$ protocols. Thus total simulation times are within the decoherence times for trapped-ion setups, of about 30 ms~\cite{Haeffner2008}.
In a digital protocol performed on real quantum systems, each gate is affected by an error. Thus, increasing the number of Trotter steps leads to an accumulation of the single gate error. To implement an effective quantum simulation, on one hand one has to increase the number of steps to reduce the error due to the digital approximation, on the other hand one is limited by the accumulation of the single gate error. We plot in Fig.~\ref{FidelityLoss}a,~\ref{FidelityLoss}b,~\ref{FidelityLoss}c, the fidelity loss $1-|\langle \Psi_S|\Psi_E\rangle|^2$ of the simulated state $|\Psi_S\rangle$ versus the exact one $|\Psi_E\rangle$, for the hydrogen Hamiltonian, starting from the initial state with two electrons in the first two orbitals. We plot, along with the digital error, three horizontal lines representing the accumulated gate error, for $n=1,2,3$ in each plot, considering a protocol with an error per Trotter step of $\epsilon=10^{-3}$ (a), $\epsilon=10^{-4}$ (b) and $\epsilon=10^{-5}$ (c). To achieve a reasonable fidelity, one has to find a number of steps that fits the simulation at a specific time. The vertical lines and arrows in the figure mark the time regions in which the error starts to be dominated by the digital error. Trapped-ion two-qubit gates are predicted to achieve in the near future fidelities of $10^{-4}$~\cite{Kirchmair09}, thus making the use of these protocols feasible. In Fig.~\ref{FidelityLoss}d we plot the behavior of the energy of the system for the initial state $|\uparrow\uparrow\downarrow\downarrow\rangle$ for the exact dynamics, versus the digitized one. Again, one can observe how the energy can be retrieved with a small error within a reduced number of digital steps.

\subsection*{Conclusions}
Summarizing, we have proposed a quantum simulation toolkit for quantum chemistry with trapped ions. This paradigm in quantum simulations has several advantages: an efficient electronic simulation, the possibility of interacting electronic and vibrational degrees of freedom, and the increasing scalability provided by trapped-ion systems. This approach for solving quantum chemistry problems aims to combine the best of classical and quantum computation. 

\section{Methods}
To implement the optimization with the UCC wavefunction ansatz on a trapped-ion quantum simulator, our proposal is to first employ classical algorithms to obtain approximate solutions~\cite{Kutzelnigg1991,Taube2006}. Then, we can further improve the quality of the solution by searching for the true minima with an ion trap. The idea is as follows: first we create a UCC ansatz by the Suzuki-Trotter method described in the previous section. Denote this choice of the cluster operator as $T^{(0)}$, and other choices as $T^{(k)}$ with $k=1,2,3,...$. The corresponding energy $E_0 = \left\langle \Phi  \right|{e^{{T^{(0) \dagger} } - T^{(0)}}}H{e^{T^{(0)} - {T^{(0) \dagger}}}}\left| \Phi  \right\rangle$ of the initial state is obtained by a classical computer. 
 
Next, we choose another set of cluster operator $T^{(1)}$ with is a perturbation around $T^{(0)}$. Define the new probe state $\left| {{\phi _k}} \right\rangle  \equiv {e^{{T^{(k)}} - T^{(k) \dagger}}}\left| \Phi  \right\rangle$. Then,
the expectation value of the energy ${E_1} = \langle \Phi |{e^{T^{(1) \dag}  - {T^{(1)}}}}H{e^{{T^{(1)}} - T^{(1) \dag} }}\left| \Phi  \right\rangle  = \left\langle {{\phi _1}} \right|H\left| {{\phi _1}} \right\rangle $ can be obtained by measuring components of the second quantized Hamiltonian, $\left\langle {{\phi _1}} \right|H\left| {{\phi _1}} \right\rangle  = \sum\nolimits_{pqrs} {{{\tilde h}_{pqrs}}\left\langle {{\phi _1}} \right|} c_p^\dagger c_q^\dagger  {c_r}{c_s}\left| {{\phi _1}} \right\rangle $. Recall that the coefficients ${{{\tilde h}_{pqrs}}}$ are all precomputed and known. 

In order to obtain measurement results for the operators $\left\langle {{\phi _1}} \right|c_p^\dagger c_q^\dagger {c_r}{c_s}\left| {{\phi _1}} \right\rangle$, we will first convert the fermion operators into spin operators via Jordan-Wigner transformation; the same procedure is applied for creating the state $\left| {{\phi _1}} \right\rangle$. The quantum measurement for the resulting products of Pauli matrices can be achieved efficiently with trapped ions, using the method we described.
 
The following steps are determined through a classical optimization algorithm. There can be many choices for such an algorithm, for example gradient descent method, Nelder-Mead method, or quasi-Newton methods. For completeness, we summarize below the application of gradient descent method to our optimization problem.

First we define the vector ${{\bf T}^{\left( k \right)}} = {( {t_i^{a\left( k \right)},t_{ij}^{ab\left( k \right)},...})^T}$ to contain all coefficients in the cluster operator $T^{(k)}$ at the $k$-th step. We can also write the expectation value $E\left( {{{\bf T}^{\left( k \right)}}} \right) \equiv \left\langle {{\phi _k}} \right|H\left| {{\phi _k}} \right\rangle$ for each step as a function of ${{\bf T}^{\left( k \right)}}$. The main idea of the gradient descent method is that $E\left( {{{\bf T}^{\left( k \right)}}} \right)$ decreases fastest along the direction of the negative gradient of $E\left( {{{\bf T}^{\left( k \right)}}} \right)$, $- \nabla E\left( {{{\bf T}^{\left( k \right)}}} \right)$. Therefore, the $(k+1)$-th step is determined by the following relation:
\begin{equation}
{{\bf T}^{\left( {k + 1} \right)}} = {{\bf T}^{\left( k \right)}} - {a_k}\nabla E( {{{\bf T}^{\left( k \right)}}}),
\end{equation}
where $a_k$ is an adjustable parameter; it can be different for each step. To obtain values of the gradient $\nabla E\left( {{{\bf T}^{\left( k \right)}}} \right)$, one may use the finite-difference method to approximate the gradient. However, numerical gradient techniques are often susceptible to numerical instability. Alternatively, we can invoke the Hellman-Feynman theorem and get, e.g., $\left( {\partial /\partial t_i^a} \right)E({{\bf{T}}^{\left( k \right)}}) = \langle {\phi _k}|[H,c_a^\dag {c_i}]\left| {{\phi _k}} \right\rangle $, which can be obtained with a method similar to that for obtaining $E( {{{\bf T}^{\left( k \right)}}} )$. 

Finally, as a valid assumption for general cases, we assume our parametrization of UCC gives a smooth function for $E\left( {{{\bf T}^{\left( k \right)}}} \right)$. Thus, it follows that $E\left( {{{\bf T}^{\left( 0 \right)}}} \right) \ge E\left( {{{\bf T}^{\left( 1 \right)}}} \right) \ge E\left( {{{\bf T}^{\left( 2 \right)}}} \right) \ge \cdots $, and eventually $E\left( {{{\bf T}^{\left( k \right)}}} \right)$ converges to a minimum value for large $k$. Finally, we can also obtain the optimized UCC quantum state.

\section*{Author Contributions}
M.-H.Y., J.M., and A.A.-G. are responsible for the parts involving quantum chemistry. J.C., A.M., L.L., and E.S. are responsible for the parts involving trapped ions. All authors contributed to the writing of the paper. 

\section*{ACKNOWLEDGMENTS}
We thank J. Whitfield for insightful discussions. The authors acknowledge funding from  Basque Government IT472-10 Grant, Spanish MINECO FIS2012-36673-C03-02, UPV/EHU UFI 11/55, SOLID, CCQED, PROMISCE and SCALEQIT European projects. M.-H.Y. and A.-A.-G. acknowledge support from the Defense Threat Reduction Agency under grant  HDTRA1-10-1-0046-DOD35CAP as well as the National Science Foundation under grant 1037992-CHE. Sponsored by United States Department of Defense.
The views and conclusions contained in this document are those of the authors and should not be interpreted as representing the official policies, either expressly or implied, of the U.S. Government.
 
\section*{ADDITIONAL INFORMATION} 
The authors declare no competing financial interests.

\bibliographystyle{apsrev}

\clearpage
\newpage

\appendix

\section{Supplementary Material}

In this Supplementary Material we give further details of our proposal, including a thorough explanation of the quantum simulation of molecules involving fermionic and bosonic degrees of freedom with trapped ions, and electric dipole transition measurements with a trapped-ion quantum simulator.

\subsection{Quantum simulation}

In general, quantum simulation can be divided into two classes, namely analog and digital. Analog quantum simulation requires the engineering of the Hamiltonian of a certain system to mimic the Hamiltonian of a target system. Digital quantum simulation employs a quantum computer, which decomposes the simulation process into pieces of sub-modules such as quantum logic gates. However, the use of quantum logic gates is not absolutely necessary for digital quantum simulation. For example, consider the case of trapped ions; we will see that certain simulation steps requires us to apply quantum logic gates to implement fermionic degrees of freedom, together with some quantum operations for controlling the vibronic degrees of freedom, which are analog and will implement bosonic modes.

For simulating quantum chemistry, it is possible to work in either the first-quantization representation or the second-quantization representation. This work mainly includes the latter approach, because the number of qubits required is less than that in the former approach, especially when low-energy state properties are considered. However, we note that many techniques described here are also applicable for the first-quantization approach.

\subsection{Computational complexity of quantum chemistry}
To the best of our knowledge, there is no rigorous proof showing that quantum computers are capable of solving all ground-state problems in quantum chemistry. Instead, some results indicate that some ground-state problems in physics and chemistry are computationally hard problems~\cite{Whitfield2012SupMat}.  For example,  the N-representability problem is known to be $\sf QMA$-complete, and finding the universal functional in density functional theory is known to be $\sf QMA$-hard. In spite of the negative results, quantum computers can still be valuable for solving a wide range of quantum chemistry problems. These include ground state energy computations~\cite{Abrams1999SupMat,Aspuru-Guzik2005cSupMat}, as well as molecular dynamics~\cite{Love2012SupMat}. 

 \subsection{Simulating electronic structure involving molecular vibrations}
After the potential surface is constructed by the electronic method, we can include the effect of molecular vibrations by local expansion, e.g. near the equilibrium position, as we show below.

\subsubsection{Electronic transitions coupled with nuclear motion}
We point out that within the Born-Oppenheimer approximation, the molecular vibronic states are of the form, ${\phi _n}\left(  {{\bf r},{\bf R}} \right){\chi _{n,v}}\left( {\bf R} \right)$ where ${\bf r}$ and ${\bf R}$ respectively refers to the electronic and nuclear coordinates. The eigenfunctions ${\phi _n}\left( {{\bf r},{\bf R}} \right)$ of the electronic Hamiltonian are obtained at a fixed nuclear configuration.  The nuclear wavefunction ${\chi _{n,v}}\left( {\bf R} \right)$, for each electronic eigenstate $n$, is defined through a nuclear potential surface $E^{(n)}_{\rm el}({\bf R})$, which is also one of the eigenenergies of the electronic Hamiltonian. 

With a quantum computer, the potential energy surface that corresponds to different electronic eigenstates can be systematically probed using the phase estimation method. We can then locate those local minima where the gradient of the energy is zero, and approximate up to second order in $\delta {R_\alpha } \equiv {R_\alpha } - {R_{\alpha *}}$, the deviation of the nuclear coordinate ${R_\alpha }$ from the equilibrium configuration ${R_{\alpha *}}$.
The energy surface can be modeled as
\begin{equation}
E_{{\rm{el}}}^{\left( n \right)}\left( {\bf R} \right) \approx E_{{\rm{el}}}^{\left( n \right)}( {{{\bf R}^{(n)}_*}} ) + \sum_{\alpha ,\beta } D_{\alpha \beta}( {{\bf R}_*^{\left( n \right)}} ) \delta {R_\alpha }\delta {R_\beta } \, ,
\end{equation}
where $D_{\alpha \beta}( {{\bf R}_*^{\left( n \right)}} ) \equiv \left( {1/2} \right){{\partial ^2}E_{el}^{\left( n \right)}( {{\bf R} {=} {\bf R}_*^{\left( n \right)}})/\partial {R_\alpha }\partial {R_\beta }}$ is the Hessian matrix. With a change of coordinates for the Hessian matrices, we can always choose to work with the normal modes ${{\bf x}^{(n)}}=\{ x_\alpha ^{\left( n \right)} \}$ for each potential energy surface, such that
\begin{equation}
E_{{\rm{el}}}^{\left( n \right)}( {\bf x}^{(n)} ) \approx E_{{\rm{el}}}^{\left( n \right)}( {{{\bf R}^{(n)}_*}}) + {1 \over 2}\sum\limits_\alpha  {{m_\alpha }\omega _\alpha ^{\left( n \right)2}x_\alpha ^{\left( n \right)2}}.
\end{equation}

Most of the important features of vibronic coupling can be captured by considering the transition between two Born-Oppenheimer electronic levels~\cite{may2011chargeSupMat}. In the following, we will focus on the method of simulation of the transition between two electronic levels, labeled as $\left|  \uparrow  \right\rangle $ and $\left|  \downarrow  \right\rangle $, when perturbed by an external laser field.  The Hamiltonian of the system 
can be written as 

\begin{equation}\label{H_HG_HE}
H = \left|  \downarrow  \right\rangle \left\langle  \downarrow  \right| \otimes {H_G} + \left|  \uparrow  \right\rangle \left\langle  \uparrow  \right| \otimes {H_E},
\end{equation}
where ${H_G} \equiv {\Delta_g} + {H_g}$ is the Hamiltonian for the nuclear motion in the electronic ground state and similarly ${H_E} \equiv {\Delta_e} + {H_e}$ is the nuclear Hamiltonian in the excited state. Here $\Delta_g$ and $\Delta_e$ are the energies of the two bare electronic states. In the second-quantized representation, 
\begin{equation}\label{HgHe}
{H_g} = \sum\limits_k {\omega _k^{\left( g \right)}a_k^ \dagger {a_k}} \quad {\rm and} \quad {H_e} = \sum\limits_k {\omega _k^{\left( e \right)}b_k^\dagger {b_k}}
\end{equation}
are diagonal, as viewed from their own coordinate systems. However, in general, the two sets of normal modes are related by rotation and translation, which means that a transformation of the kind ${b_k} = \sum\nolimits_j {{s_{kj}}{a_j} + {\lambda_k}}$ is needed for unifying the representations (see Secs.~\ref{SectAppendSpinBoson} and~\ref{AppendixMultimodeVibronic} in this Supplementary Material).

To illustrate our method of quantum simulation with trapped ions, it is sufficient to consider one normal mode (for example, linear molecules). For this case, we assume ${H_g} = {\omega ^{\left( g \right)}}{a^\dagger }a$, ${H_e} = {\omega ^{\left( e \right)}}{b^ \dagger }b$, and $b = a + \lambda$ where $\lambda$ is a real constant. From Eq.~(\ref{H_HG_HE}), we need to simulate the following Hamiltonian,
\begin{equation}\label{nucmot}
H = {H_S} + \Omega \left( {{\sigma _z}} \right)  {a^ \dag }a + {\textstyle{1 \over 2}}\lambda {\omega ^{\left( e \right)}}\left( {I + {\sigma _z}} \right) \left( {{a^ \dag } + a} \right),
\end{equation}
where the term ${H_S} = {\textstyle{1 \over 2}}\left( {{\Delta _g} - {\Delta _e}} \right){\sigma _z}$ contains only local terms of the spin, and $\Omega \left( {{\sigma _z}} \right) = {\textstyle{1 \over 2}}\left( {{\omega ^{\left( g \right)}} + {\omega ^{\left( e \right)}}} \right)I + {\textstyle{1 \over 2}}\left( {{\omega ^{\left( g \right)}} - {\omega ^{\left( e \right)}}} \right){\sigma _z}$ represents a spin-dependent frequency for the effective boson mode.

In order to examine the response of the system under external pertubations, we consider the dipole correlation
function
\begin{equation}
{C_{\mu \mu }}\left( t \right) = \sum\limits_n {{p_n}\left\langle n, \downarrow \right|{e^{iHt}}\mu } {e^{ - iHt}}\mu \left| n, \downarrow \right\rangle.
\end{equation}

Under the Condon approximation, assuming real electronic eigenstates, the dipole operator $\mu$ has the form, 
\begin{equation}
  \mu  = {\mu _{ge}} \left( \left|  \downarrow  \right\rangle \left\langle  \uparrow  \right| + \left|  \uparrow  \right\rangle \left\langle  \downarrow  \right| \right) = \mu_{ge} \sigma_x.
\end{equation}
Thus, the problem of simulating absorption resulting from the coupling of electronic and nuclear motion in chemistry reduces to computing expectation values of the unitary operator
\begin{equation}
U_d = e^{iHt} \sigma_x e^{ - iHt} \sigma_x,
\end{equation}
and weighting the final result by $p_n \mu_{ge}^2$.  The final spectrum is, of course, obtained through
a Fourier transform
\begin{equation}
  \sigma_{abs}(\omega) = \int_{-\infty}^{\infty} dt \ e^{-i \omega t} C_{\mu \mu} (t).
\end{equation}

\subsubsection{Simulation of vibronic coupling with trapped ions}
The dynamics associated with the Hamiltonian in Eq.~(\ref{nucmot}) can be generated easily with two trapped ions. As  $H_S$ commutes with the rest of the terms in Eq.~(\ref{nucmot}), it can be eliminated via a change to an interaction picture. Considering a digital quantum simulation protocol,  the remaining task  is to implement the interactions $\exp[-i\Omega\left( {{\sigma _z}} \right) t {a^ \dag }a   ]$ and $\exp[-i \lambda \omega^{(e)}\left( {I + {\sigma _z}} \right) \left( {{a^ \dag } + a} \right)  t/2]$ in trapped ions. The first one corresponds to the evolution associated with a detuned red sideband excitation applied to one of the ions (a dispersive  Jaynes-Cummings interaction), and a rotation of its internal state in order to eliminate the residual projective term.  To implement the second term we will use both ions. The term related to the operator $\sigma_z (a^{\dag} +a)$ corresponds to the evolution under red and blue sideband excitations applied to one of the ions (a Jaynes-Cummings and anti Jaynes-Cummings interactions with appropriate phases). We will use the second ion to implement the  term $(a^\dag + a)$. The latter can be generated by applying again the same scheme of lasers that generates the interaction $\sigma_z(a^\dag + a)$ where now the operator $\sigma_z$ acts on the internal state of the second ion.   Preparing this state in an eigenstate of $\sigma_z$ one obtains the desired effective Hamiltonian. As we have shown here, one of the main appeals of a quantum simulation of quantum chemistry with trapped ions is the possibility to include fermionic (electronic) as well as bosonic (vibronic) degrees of freedom, in a new kind of mixed digital-analog quantum simulator. The availability of the motional degrees of freedom in trapped ions, that straightforwardly provide the bosonic modes in an analog way, makes this system especially suited for simulating this kind of chemical problems.

\subsection{Electric transition dipoles through weak measurement}
Here we sketch the method for obtaining the transition dipole between a pair of electronic states $\left| g \right\rangle$ and $\left| e \right\rangle$. This method is similar, although not identical, to the weak measurement method using a qubit as a measurement probe. To make the presentation of our method more general, our goal is to measure the matrix element $\left\langle e \right|A\left| g \right\rangle$ for any given Hermitian matrix $A$. We assume that a potential energy surface between these two electronic levels is probably scanned, and the energy levels for higher excited states can be ignored. Suppose we started with a reasonable good approximation of the ground state $\left| g \right\rangle$, and we can prepare the exact ground state using the phase estimation algorithm. Then, we apply a weak perturbation $\lambda$, e.g. ${e^{ - i\lambda Q}}$, to the ground state and obtain (to order $O(\lambda)$) the state $\left| i \right\rangle  \equiv {e^{ - i\lambda Q}}\left| g \right\rangle  \approx \left| g \right\rangle  + q\lambda \left| e \right\rangle $. Here $\lambda$ is a small positive real number. The actual form of the Hermitian operator $Q$ is not important, as long as $\left\langle e \right|Q\left| g \right\rangle  \equiv iq \ne 0$. Note that the eigenstates are defined up a phase factor. Therefore, without loss of generality, we can assume $q$ is a positive real number as well. In fact, the absolute value $\left| q \right|$ can be measured with repeated applications of the phase estimation algorithm.

Now, we prepare an ancilla qubit in the state $\left|  +  \right\rangle  \equiv \left( {\left| 0 \right\rangle  + \left| 1 \right\rangle } \right)/\sqrt 2$, and apply a control-$U_A$, where ${U_A} \equiv {e^{ - i \lambda A}}$. The resulting state becomes $\left( {\left| 0 \right\rangle \left| i \right\rangle  + \left| 1 \right\rangle {U_A}\left| i \right\rangle } \right)/\sqrt 2$. The phase estimation algorithm allows us to perform post-selection to project the system state to $\left| e \right\rangle$. The resulting state of the ancilla qubit is $ \propto \langle e | i \rangle | 0 \rangle  + \langle e |{U_A}| i \rangle | 1 \rangle $. To the first-order expansion in $\lambda$, we have (before normalization)
\begin{equation}
q\lambda \left| 0 \right\rangle  + \left( {q - i\left\langle e \right|A\left| g \right\rangle } \right)\lambda \left| 1 \right\rangle,
\end{equation}
where we used $\langle e | i \rangle  = q\lambda$, and $\left\langle e \right|{U_A}\left| i \right\rangle  = \left\langle e \right|{U_A}\left| g \right\rangle  + q\lambda \left\langle e \right|{U_A}\left| e \right\rangle  =  - i\left\langle e \right|A\left| g \right\rangle \lambda  + q\lambda$. Since the value of $q$ is known, a state tomography on the ancilla qubit state reveals the value of the matrix element $\left\langle e \right|A\left| g \right\rangle$.
 
Returning to the case of the electric dipole moment, it is defined as ${\mathbf{\mu}}  \equiv  - e\sum\nolimits_i {{{\bf r}_i}} $. In the second quantized form is $\mu  = \sum\nolimits_{pq} {{u_{pq}}a_p^\dag {a_q}}$, where ${u_{pq}} \equiv  - e\int {\phi _p^*\left( {\bf{r}} \right)} {\bf{r}}{\phi _q}\left( {\bf{r}} \right)d{\bf{r}}$ is nothing but the single-particle integral. The dipole operator therefore has the same form as the first term in Eq.~(\ref{h_pq_h_pqrs}), and the simulation of the corresponding operator ${U_A} \equiv {e^{ - i \lambda A}}$, with $A$ replaced by $\mu$, can be simulated efficiently after performing the Jordan-Wigner transformation.
 
\subsection{Derivation of the spin-boson coupling\label{SectAppendSpinBoson}}
Consider the full Hamiltonian of two potential energy surfaces,
\begin{equation}
H = \left|  \downarrow  \right\rangle \left\langle  \downarrow  \right| \otimes {H_G} + \left|  \uparrow  \right\rangle \left\langle  \uparrow  \right| \otimes {H_E},
\end{equation}
where 
\begin{equation}
{H_G} \equiv {\Delta_g} + {H_g}
\end{equation}
is the Hamiltonian for the nuclear motion in the electronic ground state and similarly 
\begin{equation}
{H_E} \equiv {\Delta_e} + {H_e}
\end{equation}
is the nuclear Hamiltonian in the excited state. Here $\Delta_g$ and $\Delta_e$ are the zero-point energies of the two potential energy surfaces. In the second-quantized representation, we consider one normal mode for each local minimum in the potential energy surface,
\begin{equation}
{H_g} = {\omega ^{\left( g \right)}}{a^\dagger }a \quad {\rm and} \quad {H_e} = {\omega ^{\left( e \right)}}{b^ \dagger }b.
\end{equation}
Here the two normal modes are related by a shift of a real constant $\lambda$, namely
\begin{equation}\label{app:b=a+L}
b = a + \lambda.
\end{equation}
Now, we will rewrite the full Hamiltonian in terms of the Pauli matrix 
\begin{equation}
{\sigma _z} = \left( {\begin{array}{*{20}{c}}
1&0\\
0&{ - 1}
\end{array}} \right) = \left|  \uparrow  \right\rangle \left\langle  \uparrow  \right| - \left|  \downarrow  \right\rangle \left\langle  \downarrow  \right|.
\end{equation}
First of all, we write $H = {H_{SB}} + {H_S}$, where 
\begin{equation}
H_{SB} = \left|  \downarrow  \right\rangle \left\langle  \downarrow  \right| \otimes {\omega ^{\left( g \right)}}{a^\dagger }a + \left|  \uparrow  \right\rangle \left\langle  \uparrow  \right| \otimes {\omega ^{\left( e \right)}}{b^\dagger }b,
\end{equation}
and 
\begin{eqnarray}
{H_S} &\equiv  & \left|  \downarrow  \right\rangle \langle  \downarrow |{\Delta _g} + \left|  \uparrow  \right\rangle \langle  \uparrow |{\Delta _e} \nonumber \\
& =  & \frac{1}{2}\left( {{\Delta _g} + {\Delta _e}} \right)I + \frac{1}{2}\left( {{\Delta _g} - {\Delta _e}} \right){\sigma _z}.
\end{eqnarray}

Next, we use Eq.~(\ref{app:b=a+L}) to write $H_{SB}$ as
\begin{equation}
H_{SB}=\Omega \left( {{\sigma _z}} \right) \otimes {a^ \dag }a + {\textstyle{1 \over 2}}\lambda {\omega ^{\left( e \right)}}\left( {I + {\sigma _z}} \right) \otimes \left( {{a^ \dag} + a} \right),
\end{equation}
where the frequency of the effective mode becomes spin-dependent, 
\begin{eqnarray}
\Omega \left( {{\sigma _z}} \right) & \equiv  & \left|  \uparrow  \right\rangle \langle  \uparrow |{\omega ^{\left( g \right)}} + \left|  \downarrow  \right\rangle \langle  \downarrow |{\omega ^{\left( e \right)}}\\
 &=  & \frac{1}{2}\left( {{\omega ^{\left( g \right)}} + {\omega ^{\left( e \right)}}} \right)I + \frac{1}{2}\left( {{\omega ^{\left( g \right)}} - {\omega ^{\left( e \right)}}} \right){\sigma _z} \nonumber.
\end{eqnarray}

\subsection{Multimode extension of simulating vibronic coupling\label{AppendixMultimodeVibronic}}
In order to extend the method of simulating vibronic coupling to the case with multiple bosonic modes, we now consider the case of Eq. \ref{HgHe}.  If we express the excited state modes in terms of the ground state modes such that 
\begin{equation}
b_k = \sum_j s_{kj} a_j + \lambda_k,
\end{equation}
we can write $H$ as
\begin{equation}
H = H_s' + \left|  \downarrow  \right\rangle \left\langle  \downarrow  \right| \otimes {H_G} + \left|  \uparrow  \right\rangle \left\langle  \uparrow  \right| \otimes {H_E},
\end{equation}
where
\begin{equation}
{H_G} \equiv \sum\limits_k {\omega _k^{\left( g \right)}a_k^\dagger {a_k}} ,
\end{equation}
and 
\begin{equation}
{H_E} \equiv \sum\limits_{kjl} {\omega _k^{\left( e \right)}{s_{kj}}{s_{lk}}a_j^\dag {a_l}}  + \sum\limits_{kj} {\omega _k^{\left( e \right)}} {s_{kj}}{\lambda _k}\left( {a_j^\dag  + {a_j}} \right) \, .
\end{equation}
In the definition of $H_s'$, the only change from $H_s$ is given by
\begin{equation}
\Delta_e' = \Delta_e + \sum_k \lambda_k^2 .
\end{equation}
With knowledge of $s_{ij}$ and $\lambda_i$ for all modes, we
can then repeat the above procedure to determine the absorption spetrum for a complicated system using
a quantum computer.  The above Hamiltonian can be written in a form more familiar to quantum computation as
\begin{align}
  H &= H_s' + \sum_k \Omega_k(\sigma_z) a_k^\dagger a_k \notag \\
  &+ \frac{1}{2} \sum_{kj} s_{kj} \omega_k^{(e)}\lambda_k(I + \sigma_z)(a_j^\dagger + a_j) \notag \\
  &+ \frac{1}{2} \sum_k \sum_{j \neq l} s_{kj} s_{lk} \omega_k^{(e)}(I + \sigma_z) a_j^\dagger a_l
\end{align}
where we define 
\begin{equation}
\Omega_k(\sigma_z) \equiv \frac{1}{2}(\omega^{(g)}_k + s_{kk}^2 \omega^{(e)}_k) I 
+ \frac{1}{2}(\omega^{(g)}_k - s_{kk}^2 \omega^{(e)}_k) \sigma_z.
\end{equation}
In cases where Duchinsky rotations of the normal modes can be neglected ($s_{ij} = \delta_{ij})$, this expression
can be further reduced to
\begin{align}
  H &= H_s' + \sum_k \Omega_k'(\sigma_z) a_k^\dagger a_k   \notag \\
  &+ \frac{1}{2} \sum_{k} \omega_k^{(e)}\lambda_k(I + \sigma_z)(a_k^\dagger + a_k)
\end{align}
with the simplification 
\begin{equation}
\Omega_k'(\sigma_z) = \frac{1}{2}(\omega^{(g)}_k + \omega^{(e)}_k) I 
+ \frac{1}{2}(\omega^{(g)}_k - \omega^{(e)}_k) \sigma_z.
\end{equation}

\bibliographystyle{unsrt}

\end{document}